# The impact of introductory physics for the life sciences in a senior biology capstone course

Benjamin D. Geller,[1] Jack Rubien,[1] Sara M. Hiebert,[2] and Catherine H. Crouch[1]

[1]*Department of Physics & Astronomy, Swarthmore College, Swarthmore, PA 19081*
[2]*Department of Biology, Swarthmore College, Swarthmore, PA 19081*

## Abstract

A goal of Introductory Physics for Life Sciences (IPLS) curricula is to prepare students to effectively use physical models and quantitative reasoning in biological and medical settings. To assess whether this goal is being met, we conducted a longitudinal study of the impact of IPLS on student work in later biology and chemistry courses. We report here on one part of that study, a comparison of written responses by students with different physics backgrounds on a diffusion task administered in a senior biology capstone course. We observed differences in student reasoning that were associated with prior or concurrent enrollment in IPLS. In particular, we found that IPLS students were more likely than non-IPLS students to reason quantitatively and mechanistically about diffusive phenomena, and to successfully coordinate between multiple representations of diffusive processes, even up to two years after taking the IPLS course. Finally, we describe methodological challenges encountered in both this task and other tasks used in our longitudinal study.

## I. INTRODUCTION: ASSESSING THE LONG-TERM OUTCOMES OF INTRODUCTORY PHYSICS FOR LIFE SCIENCES

In response to national calls to better train future physicians, biologists, and medical researchers in physics [1–4], many instructors have designed and delivered introductory physics curricula that are specifically aimed at addressing the needs of life science students [5–7]. These Introductory Physics for Life Sciences (IPLS) curricula emphasize quantitative reasoning and physical and computational modeling skills that will be required of life science students as they move on to careers in clinical research and medicine [7–10]. Much work has been done to assess attitudinal and skill-based outcomes *within* IPLS courses. At Swarthmore College, we have observed increased student interest and engagement, particularly among students entering the course with initially low levels of interest in physics [11]; we also find that the life science examples that provide the framework for the course seem to produce a learning environment in which physics is seen as providing a meaningful and complementary explanation for biological phenomena familiar to students [12]. However, little work to date has explored the lasting impact of these courses after students have left the IPLS environment.

Here we report findings of one element of our pilot longitudinal study of the long-term durability of IPLS outcomes, in which we studied the work of students with various physics backgrounds on a task delivered in a biology capstone course. Performance on this task represents a measure of students' ability to use physical and quantitative reasoning to analyze an authentic biological problem that is presented in a biology classroom rather than in the IPLS environment.

We found that IPLS students were more likely than non-IPLS students to provide a mechanistic description of diffusion, and did so up to two years after taking the IPLS course. We also found that IPLS students were more likely than non-IPLS students to reason quantitatively about diffusive phenomena and to successfully coordinate between multiple representations of diffusive processes. These findings provide an encouraging indication about the durability and impact of core competencies that are emphasized in the IPLS curriculum.

Section II describes the IPLS course design and theoretical framework. Section III describes the student populations in our study, as well as features of the IPLS and biology capstone courses; Section IV describes the conception, design, and administration of the diffusion task and Section V its analysis; Section VI presents the results and our interpretation of those results. Finally, Section VII offers open questions and next steps.

Longitudinal assessment of student learning is complex, and interdisciplinary longitudinal work is even more so. Interdisciplinary learning is shaped by students' epistemological expectations about each discipline, which have been shown to make transfer between disciplines challenging [13–16]; instructors of undergraduate science courses frequently communicate these disciplinary epistemologies implicitly or explicitly [7,8]. Life science students' personal identities as disciplinary scientists have been shown to shape how they view their work and capability in different disciplines [17], even when environments have been designed specifically to facilitate cross-disciplinary understanding [10,18–21]. In discussing the conception of the capstone task whose results we report (Section IV), we also discuss methodological lessons learned from prior attempts to study interdisciplinary learning.

## II. IPLS Course Design and Theoretical Framework

The Swarthmore IPLS curriculum is built around *authentic* biological contexts that are drawn directly from examples that students encounter in their life science classes [14,15]. Authenticity refers here to the curricular goal that *students* would perceive the inclusion of biological contexts to be integral to the narrative of the course, rather than to be an attempt to find biological meaning where none actually exists. We do not, for example, consider the replacement of a car with an animal in a standard kinematics problem to be authentic; we do, however, spend considerable time in IPLS 2 using ideas about electric potential and simple circuits to model the behavior of cell membranes and neural signaling. These biological contexts are not designed as optional "add-on" applications to be tackled only after the core physical ideas were learned in a traditional way; rather, they were integral to the course and repeatedly referred to throughout each unit as the physical ideas were developed. Throughout the course, emphasis is placed not only on modeling complex biological situations, but also on developing a mechanistic understanding of biological and biophysical phenomena.

Two pedagogical strategies undergirding the Swarthmore IPLS course complement the curricular focus on authentic biological contexts: expansive framing [22–24] and cognitive apprenticeship [25]. Expansive framing refers to the goal of presenting the conceptual material in a way that allows students to be see that it is broadly applicable not only to the scientific community outside the physics classroom but also to students' future interests and careers [24]. IPLS course content is explicitly framed as being relevant and connected to students' other coursework in biology and chemistry, both now and in the future. In exploring these connections, the instructor encourages students to draw on their own backgrounds in biology and chemistry, and explicitly positions students as having expertise in areas with which the instructor might have little familiarity.

The other pedagogical strategy essential for supporting students in applying what they learn in later work is called "cognitive apprenticeship" [25]. Cognitive apprenticeship shares elements with the "communities of practice" approach [26]. In a cognitive apprenticeship framework, the goal is to create a learning environment that has essential features in common with the environment in which an expert functions. Specifically, such an environment repeatedly prompts the apprentice to assess (i) why they are learning what they are learning, and (ii) how what they are learning connects to things they already know. Within the cognitive apprenticeship framework, the classroom is meant to simulate this sort of environment as closely and as frequently as possible. In the context of the IPLS course at Swarthmore, the instructor routinely demonstrates the process of complex problem-solving as part of an interactive lecture, with particular attention given to the decision-making steps and simplifying assumptions that are essential for describing complex biological systems with simple physical models. Explicit articulation of these decision-making steps is an essential aspect of the modeling stage of the apprenticeship.

Ultimately, our IPLS course is designed to support students to use the tools and concepts of physics in their later biomedical studies and careers, and thus it ultimately has interdisciplinary transfer as a core goal. Our approach to this goal is grounded in the "preparation for future learning" concept of transfer [27,28]. We are not seeking to prepare students to be full-blown biological physicists in a single course, which seems an unrealistic goal. Rather, our course goal is to make students more able to recognize biomedical situations in which using physical ideas and strategies will add to their understanding, and will be amenable to seeking the resources needed to do so. Prior work demonstrates that expansive framing strategies can support this type of transfer [24].

## III. STUDENT GROUPS AND COURSE CONTEXTS

### A. The Biology Capstone Course

In this study, we analyzed written responses to a diffusion task from a total of 64 biology majors enrolled in a biology capstone course. The capstone course is required for biology majors at Swarthmore, and is typically taken by seniors. It is designed to be a culminating experience for biology majors, and most course meetings involve interaction with primary literature and expert guest speakers. Throughout the course, students are expected to synthesize what they have learned during their time as a biology major at Swarthmore.

While physics can be used to satisfy the biology major's requirement for coursework in related sciences (called the "quantitative course requirement" because all of the permitted courses are purported to involve using quantitative skills), this requirement can be satisfied by taking courses other than physics. For this reason, 17 of the 64 students in our sample had not taken physics at Swarthmore at all. Although life science students at Swarthmore typically take the IPLS course if they take a physics course, during the time of this study, the first semester of IPLS ("IPLS 1") was offered only in alternate years because of staffing limitations. Thus, an additional 9 students in our study took non-IPLS college physics. Finally, students take the IPLS course at various times during their undergraduate career, and sometimes take the two semesters out of order, allowing us to probe the temporal durability of skills developed in IPLS. A table summarizing the number of subjects with each type of physics background, including the time between taking IPLS 1 (the IPLS semester most relevant to the diffusion task) and the capstone course, is presented with the methodology in Section V.

### B. The IPLS Course

Swarthmore offers a two-semester IPLS sequence: IPLS 1 (Mechanics) and IPLS 2 (Electricity, Magnetism, and Optics). The IPLS course is nominally calculus-based, but in practice very little calculus is used and students are rarely asked to evaluate integrals. Although there are no formal biology or chemistry prerequisites for the IPLS course, the vast majority of students in the course have taken or are co-enrolled in both biology and chemistry coursework. Enrollment in IPLS is typically about 50 students, with each class consisting of approximately 10% first-year students, 75% sophomores and juniors, and 15% seniors.

The IPLS course is lecture-based, but think-pair-share tasks and opportunities for group problem solving are used throughout most lectures, and discussion among students is encouraged by instructors and facilitated by undergraduate peer assistants. IPLS students are expected to engage with biological contexts in multiple ways: through activities during the interactive lecture, through problems on homework sets that go beyond the basic ideas presented in lecture, and through context-rich scenarios posed during recitation and lab sessions. Models are often developed in an iterative fashion as the course progresses. For example, the model of the electrical properties of a cell membrane is developed gradually as new electrical ideas (resistance, capacitance, current) are encountered. In this way, each time a new physical property is described, students immediately encounter its relevance for modeling a real biological system. We view this process as essential for supporting students in using physics after leaving the IPLS environment.

Diffusion, the context for the task analyzed in this study, is discussed in IPLS 1 as part of the study of random motion. Particular emphasis is placed on understanding how, although *individual* molecules move entirely randomly, the spontaneous *net movement* of a chemical species proceeds from regions of high concentration to regions of low concentration. The mechanism is subtle. While students know the heuristic that molecules move from regions of high to low concentration, they frequently struggle to reconcile these microscopic and macroscopic ideas as they are learning about diffusive flow in the IPLS 1 course. In contrast, random motion and diffusive flow are not discussed in detail in the non-IPLS Physics 1 course.

## IV. DIFFUSION TASK CONCEPTION, DESIGN, AND DELIVERY

### A. Strategy for Designing the Diffusion Task

Design of the biology capstone diffusion task followed two years of previous efforts to assess the long-term outcomes of our IPLS courses at Swarthmore. While initial findings from those prior analyses suggested that improved *attitudes* about the relevance of physics to the life sciences last at least two years after students finish the courses [29], preliminary efforts to assess the durability of *skills* developed in IPLS proved more challenging, and revealed complexities inherent in cross-disciplinary longitudinal work [30–32]. Our initial approach to longitudinal assessment of IPLS skills was to collect and analyze student written work on "embedded tasks" given by the instructors of intermediate and advanced biology and chemistry courses that students took after they had completed introductory physics. For this initial approach, the physicist IPLS researchers (BG and CHC) had little or no input into the design of the embedded task prompts to which students were responding; we asked course instructors to teach their classes exactly as they normally would.

While this initial hands-off approach had the potential to reveal robust evidence of unprompted quantitative and physical reasoning across disciplinary boundaries, this kind of reasoning proved difficult to discern for several reasons. Questions about whether and when it was appropriate to combine data related to common skills but collected across different biology course environments made building a sufficient sample size challenging. It was also difficult to disentangle learning gained in the IPLS environment from learning gained within the intermediate and advanced biology courses themselves. While those biology courses rarely emphasized physical models in detail, the course instructors did on occasion reference such models, making it hard to know whether any physical reasoning we observed was due to IPLS or to learning that happened in the biology courses themselves.

Most importantly, because the embedded task prompts rarely explicitly required quantitative responses, even students who might possess significant quantitative or physical reasoning skills might not have thought it was appropriate to demonstrate those skills in that context. For students to exhibit unprompted quantitative and physical reasoning on tasks embedded in later biology or chemistry coursework, they must realize the value of using those skills in contexts that do not explicitly call for them. In biology courses that do not frequently emphasize physical or quantitative reasoning, for example, students may not see such reasoning as relevant, even if they can in fact draw on those skills when explicitly asked to do so. We developed emergent coding schemes to identify unprompted physical reasoning, but the absence of such reasoning did not conclusively indicate that the student lacked the relevant skill.

The results from these embedded tasks for which we as researchers had little or no input were encouraging,[1] but inconclusive. Larger sample sizes may eventually result in statistical significance for some of the preliminary differences we observe, but the challenges described above made it clear that detecting a more robust signal required us as researchers to ask more directly about desired competencies.

The biology capstone task used these more direct prompts. Part I of the diffusion task, for example, includes a prompt explicitly asking students to describe the mechanism underlying diffusive flow in the context of animal digestion. This prompt is included because prior efforts to measure mechanistic reasoning about tasks in which students were not explicitly prompted for a physical mechanism were inconclusive. Similarly, other prompts within the diffusion task call more explicitly for physical and quantitative reasoning than we had done in our prior work on embedded task analysis, while still maintaining the authenticity of the biological context in which the task is situated.

---

[1] On every embedded task for which we developed emergent codes, the median performance of IPLS students was similar or better than that of their peers who did not take IPLS.

## B. Designing the Diffusion Task

SMH (a biologist) and BG (a physicist) closely collaborated on the design of the diffusion task (Appendix). Diffusion was chosen as an ideal context because all biology students encounter diffusive or "gradient-driven" flow in one or more of their biology courses at Swarthmore, and because diffusion is discussed in some mechanistic detail in IPLS 1. SMH suggested the specific biological context for the diffusion task, the diffusion of fatty acid molecules from the lumen of the small intestine to a blood vessel in the intestinal wall. This context had the advantage that it is one that students *do not* encounter in IPLS 1 (or in any biology course at Swarthmore), and yet can be analyzed in the same way as other diffusive contexts that *are* discussed in IPLS 1. BG was the primary writer of the detailed prompts, which were informed by his teaching of diffusion in IPLS 1 and by his understanding of the quantitative skills that students have the opportunity to develop in both IPLS 1 and the other courses that meet the biology department's quantitative course requirements for majors. SMH ensured that the biological context and content contained in the prompts was appropriate.

The diffusion task consists of three connected parts, each of which assesses different physical and quantitative competencies (Table I):

| Diffusion task element | Physical or quantitative competencies assessed |
|---|---|
| **Part I** | - To *produce a qualitatively accurate graph* (of fatty acid concentration as a function of position) from a written description of the biological phenomenon<br>- To *describe the mechanism* of diffusive (gradient-driven) flow at a molecular level. |
| **Part II** | - To *interpret quantitative information* about diffusion that is presented graphically.<br>- To *compare slopes quantitatively* in order to determine relative rates of diffusion. |
| **Part III** | - To *interpret the meaning of a quantitative relationship* (Fick's Law) that is provided in the problem statement.<br>- To *calculate rates of diffusion* from numerical and graphical data, and to *explain the results of these calculations conceptually*. |

**Table I.** The physical or quantitative competencies assessed in each part of the diffusion task. The full task is provided as an Appendix.

Part I was designed to probe students' abilities (a) to use a provided written description of a diffusive phenomenon to sketch a graph that represents molecular concentration as a function of position, and (b) to articulate the molecular mechanism underlying diffusive (gradient-driven) flow. In particular, Part I probes whether students can provide a sound mechanistic explanation for why molecules diffuse from high to low concentration, even though each individual molecule moves randomly. This mechanism is a subtle idea that is discussed in IPLS 1, but rarely if ever discussed in great detail in the biology courses that students take. Life science students are almost always comfortable with the heuristic that molecules diffuse from regions of high to low concentration, but are much less comfortable reconciling this idea with the randomness of individual molecular movements.

Part II was designed to assess a student's ability to interpret quantitative information about diffusion that is presented graphically. In particular, Part II probes student understanding of the meaning of a slope as a rate and student ability to compare slopes quantitatively in order to determine relative rates of diffusion.

Part III was designed to test students' abilities to use and interpret a quantitative relationship provided in the problem statement. The introduction to Part III states Fick's Law mathematically and defines each term in the equation carefully, so that no recall is required. Students are then asked to calculate rates of diffusion from numerical and graphical data, and to explain their answers conceptually. The conceptual explanation is subtle, as it involves relating the sign of the diffusion rate to a spatial direction. This subtlety is discussed in IPLS 1, but many students nonetheless struggle to coordinate the sign with spatial direction and instead think of the sign as always referring to the direction of decreasing molecular concentration.

All three parts of the task probe students' abilities to (a) read and interpret verbal and graphical depictions of biological data, (b) coordinate between verbal descriptions and graphical representations of these data, and (c) use simple physical models to analyze complex biological situations, though to varying degrees and in different ways. These skills are emphasized and practiced repeatedly in both semesters of the IPLS course, and students are provided with opportunities to develop these skills in the other courses that biology students take to meet the department's quantitative course requirement.

### C. Administering the Diffusion Task

We administered the diffusion task during the biology capstone course in two consecutive years (33 students completed the task in the first year, and 31 the second year). Because of the COVID-19 pandemic, the second iteration of the capstone course was taught remotely, but the results obtained by analyzing student responses from the two iterations of the course were qualitatively and quantitatively similar (see Section VI).

The diffusion task was administered to students at the final meeting of the capstone course, by a biology faculty member (SMH). The task was framed for students as helping the biology department evaluate how well its quantitative course requirement for majors was working, and students were encouraged to draw on what they had learned in other Natural Science and Engineering (NSE) courses. In introducing the task, SHM told students that they were welcome to draw on ideas encountered in other science courses in order to invite students – but not require them – to bring in ideas from outside the capstone course. Because no specific mention of "physics" was made, students were not primed to think of the task as relating to physics in particular. The introduction was delivered by video during the second iteration, as the COVID-19 pandemic required the class to be taught remotely.

In the introduction, students were primed to think about the mechanism of diffusion in two ways: (a) they were shown a short animation of a single (dye) molecule bouncing around as it collides with other (water) molecules in a container [33], and (ii) they were shown a short video of a blob of many dye molecules diffusing through a beaker of water [34]. The purpose of these short demonstrations was to orient students – even those who had not taken physics – to the phenomena of diffusion that they had previously encountered. The hope was that showing an animation involving molecular collisions might prime students who had not taken physics to reason about the molecular mechanism responsible for gradient-driven flow.

Because Part II of the diffusion task provided students with graphs like the ones we asked students to sketch in Part I, we collected responses to Part I from the students before distributing Parts II and III. All parts of the task were administered on paper in the first year, but because of the COVID-19 pandemic the task was administered online in the second year and students uploaded their written work to the course website. There was no specific time limit for completing the task, but almost all students completed the entire task in 30-60 minutes.

Students were encouraged to take the task seriously, but were told they would receive full credit for thoughtfully completing the task. The task was required during the first iteration of the capstone course, but because of the challenges of remote assessment, not during the second. Instead, students in the second iteration were given a small amount of extra credit for completing the task, resulting in a completion rate of about 80%.

## V. TASK ANALYSIS METHODOLOGY
### A. Emergent Code Development

We used emergent coding to categorize students' written responses to the capstone diffusion task. The code development was a highly iterative process. BG, JR, and CHC generated an initial code; then, each applied the code independently to a subset of anonymized student responses . The three coders compared their scores, refined the code to address minor ambiguities that were revealed by the initial round of coding, and the new code was then applied to a different subset of anonymized responses. Following several such iterations, the final code was validated for inter-rater reliability (Cohen's kappa between 0.8 and 1 for all coding elements), and then applied to all anonymized student responses in the study by BG. Throughout emergent code development, students' physics backgrounds were concealed; after coding was complete, student prior course histories were unmasked for further analysis.

The code itself was divided into two main sections: (1) competencies related to content and skills specifically emphasized in IPLS 1, and (2) general quantitative skills. We divided our code in this way because, while any biology major who has satisfied the biology department quantitative course requirement might be expected to demonstrate an array of general quantitative competencies, one would not expect every biology major to demonstrate the same proficiency in content or competencies that are specifically emphasized in IPLS 1.

Table II shows the specific code elements that fell into each of these two categories. The IPLS 1-specific coding elements (labeled "IPLS 1" in Table II) related to the mechanistic and graphical descriptions of diffusion, and to the coordination between the sign of diffusive flux and a direction in space. Both of these skills are explicitly emphasized in IPLS 1. The general quantitative skill elements of the code (labeled "QUANT" in Table II) include reasoning with units, comparing slopes of graphs, and using equations. These skills are developed in IPLS, but also in many of the other courses that meet the biology department quantitative course requirements.

| Part I | |
|---|---|
| **Skill** | **Scoring** |
| *Converting a written description of a biophysical scenario into a qualitatively accurate graph* | • Ends: +0.5 for each constant (horizontal) end of the sketched line<br>• Middle **(QUANT)**: +2 points if linearly decreasing; +1 if decreasing, but not linearly<br>• For bar plot or scatter plot instead of a continuous graph: +1 if trend is correct |
| *Providing a mechanistic, molecular-level explanation for the flow of molecules down a concentration gradient* | Explanation **(IPLS 1)**:<br>• +2: Difference in number of molecules between high and low concentration regions explains the net flow of particles, even though each individual molecule moves randomly<br>• +1: Explains the flow in terms of general physical reasoning (collisions, thermodynamics, Fick's law), but doesn't employ a complete mechanistic explanation<br>• 0: Restates the question or no coherent explanation<br><br>Diagram **(IPLS 1)**:<br>• +1: Diagram demonstrates why more molecules move across a boundary from high to low concentration than from low to high concentration<br>• +0.5: Diagram is present, but doesn't clearly articulate the above idea<br>• 0: No diagram |
| Part II | |
| **Skill** | **Scoring** |
| *Calculating rates of diffusion from graphical representations of concentration as a function of position.* | Correctness **(QUANT)**:<br>• +2: Completely correct ranking: B > A = D > C.<br>• +1: Slope B is steepest and slope C is least steep, but slopes A and D are not identified as having the same slope<br>• 0: Other ranking<br><br>Slope reasoning **(QUANT)**:<br>• +2: Correct reasoning with slopes<br>• +1: Incorrect calculation or incomplete explanation with slopes<br>• +0: No evidence of reasoning with slopes |
| Part III | |
| **Skill** | **Scoring** |
| *Understanding the minus sign in Fick's Law as a mathematical representation of the idea that molecules move from areas* | • **(QUANT)** +1: The minus sign is needed to specify direction of flux |
| *of high to low concentration (down their concentration gradient).* | |
| *Converting a graphical representation of diffusion into a quantitative, symbolic representation (Fick's law) to correctly calculate a rate of flux; relating the minus sign in Fick's law to the flow of fatty acids from the lumen of the intestine to the blood vessel (i.e., coordinating the sign with a spatial direction).* | End regions **(QUANT)**: <br> • +1: Identifies the ends as $J = 0$ <br><br> Middle region **(QUANT)**: <br> • +1: Correct calculation for the middle as $J = 10{,}000$ molecules/s <br> • +1: Correct sign (positive) obtained by correct use of Fick's Law <br><br> Holistic over all of Part III **(IPLS 1)**: <br> • +2: Successfully coordinates the positive sign to the direction of flow along the *x*-axis <br> • +1: Attempts to relate the sign to the coordinate system, but unsuccessfully |

**Table II.** The emergent code for analyzing the biology capstone diffusion task. IPLS 1-specific elements of the code are designated IPLS 1, and general quantitative skill elements are designated QUANT.

### B. Comparing Student Outcomes

For elements of the code that were specifically emphasized in IPLS 1, we divided students into those who had and those who had not taken IPLS 1, to test the hypothesis that IPLS 1 students would score higher than their non-IPLS 1 peers on the IPLS 1-specific elements of the code. Because of the small sample sizes and non-normally distributed data (as verified by the Anderson-Darling normality test) we compared the score distributions from the two student groups using a Mann-Whitney-Wilcoxon test (also known as a Wilcoxon two-sample test, hereafter referred to as a "Wilcoxon" test). To correct for multiple comparisons, we applied a Bonferroni correction [35,36] to the Wilcoxon test results as follows: The critical value (alpha) was divided by the number of pairwise comparisons to determine the Bonferroni-corrected critical value (e.g., for three pairwise comparisons, $0.05/3 = 0.0167$, and only those p-values < 0.0167 were interpreted as statistically significant). The Bonferroni correction is known to be highly conservative [37].

We report results obtained using non-parametric statistical tests throughout this article, since the data were found to be nonnormal. However, we also ran tests that assume normal distributions (t-tests and ANOVA/ANCOVA) and obtained qualitatively and quantitatively similar results. This finding is consistent with ambiguity in the literature about when it is necessary to use non-parametric tests. To assess whether it was appropriate to combine data obtained from the two different years in which we administered the diffusion task in the capstone course, we used Wilcoxon tests to compare (i) the results of the IPLS 1 students from each year on the IPLS 1-specific elements of the code, and (ii) the results of the non-IPLS 1 students from each year on the IPLS 1-specific elements of the code. To control for overall academic performance, an ANCOVA was performed using average science and engineering course GPA as a covariate.

To assess whether a non-IPLS 1 physics course improved student performance on IPLS 1-specific elements of the code, and to assess whether it mattered how recently a student had taken the IPLS 1 course, the IPLS 1 and no IPLS samples were further subdivided into five groups (Table III): (i) those who had taken no college physics at all, (ii) those who had taken non-IPLS college physics, (iii) those who took IPLS 1 concurrently with the biology capstone course, (iv) those who had taken IPLS 1 one year before the biology capstone course, and (v) those who had taken IPLS 1 two or more years before the biology capstone course. The IPLS 1-specific scores for these subgroups were compared using the Kruskal-Wallis test with Wilcoxon post-hoc for pair-wise comparisons.

| Capstone Study Population (N = 64) | | | | | |
|---|---|---|---|---|---|
| **No IPLS (N = 26)** | | **Some IPLS (N = 38)** | | | |
| No college physics (N = 17) | Non-IPLS college physics (N = 9) | IPLS 1 (N = 28) | | | IPLS 2 only (N = 10*) |
| | | Concurrent with capstone (N = 9) | One year before capstone (N = 9) | Two or more years before capstone (N = 10) | * 17 of IPLS 1 students also took IPLS 2. |

**Table III.** Physics backgrounds of students in the capstone study. Students in the capstone course either took no IPLS physics (blue) or took at least some IPLS physics (green). Among those who did not take IPLS physics, some took no college physics and some took non-IPLS college physics. Among those who did take at least some IPLS physics, some took only one semester of IPLS and others took two. The students who took IPLS 1 did so at different time points relative to their enrollment in the capstone course.

To assess whether IPLS students scored higher on general quantitative skill elements of the code, we divided the entire capstone population into two broad groups: (i) students who had not taken an IPLS course, and (ii) students who had taken at least one semester of IPLS. Because of the small sample sizes and nonnormally distributed data (as verified by the Anderson-Darling normality test), we used a Wilcoxon test to compare the score distributions from these two groups of students. To assess whether it was appropriate to combine data obtained from the two different years in which we administered the diffusion task in the capstone course, we used Wilcoxon tests to compare (i) the results of the IPLS students from each year on the general quantitative skill elements of the code, and (ii) the results of the non-IPLS students from each year on the general quantitative skill elements of code. Again, an ANCOVA test was used to control for overall student performance using average science and engineering GPA as a covariate. To assess whether the development of quantitative skills might be cumulative, the IPLS student group was further subdivided into students who had taken just one semester of IPLS and students who had taken both semesters of IPLS. We compared the scores of these groups on the quantitative skill elements of the code using the Kruskal-Wallis test with Wilcoxon post-hoc pair-wise comparisons.

Finally, to test the hypothesis that performance on IPLS 1 code elements does in fact depend on content and skills developed specifically in IPLS 1, while performance on general quantitative skill code elements does not depend on the specific content covered in either semester, we used a Wilcoxon test to compare performance on the various code elements between those who had taken IPLS 1 and those who had taken only IPLS 2.

All statistical analyses were performed in the R software environment.

### VI. RESULTS AND DISCUSSION

For all results reported in this section, data are combined from the two years in which the diffusion task was administered in the capstone course. On the elements of the code that directly relate to content and skills emphasized in IPLS 1, we found no statistically significant difference between the results of IPLS 1 students completing the task in each year (Wilcoxon $p = 0.75$) or between the results of non-IPLS 1 students in each year (Wilcoxon $p = 0.37$). Likewise, on the elements of the code that related to general quantitative skills, we found no statistically significant difference between the results of IPLS students completing the task in each year (Wilcoxon $p = 0.25$), or between the non-IPLS students in each year (Wilcoxon $p = 0.38$).

### A. Content and Skills Specific to IPLS 1

On elements of the code that relate directly to content and skills emphasized in IPLS 1, IPLS 1 students scored significantly higher than non-IPLS 1 students (Wilcoxon $p < 0.0001$, Fig. 1a). This difference remains significant when those who did not take IPLS 1 are compared with each of three different subgroups of the overall IPLS 1 student group (Kruskal-Wallis $p = 2.5 \times 10^{-5}$, Fig. 1b): (i) those completing the capstone task concurrently with their enrollment in IPLS 1 (Wilcoxon $p = 0.006$), (ii) those completing the capstone task one year after completing IPLS 1 (Wilcoxon $p = 0.004$), and (iii) those completing the capstone task two or more years after completing IPLS 1 (Wilcoxon $p = 0.024$). However, the difference between those who did not take IPLS 1 and those completing the capstone task two or more years after completing IPLS 1 did not survive the Bonferroni correction (Bonferroni corrected alpha = 0.0167). For the IPLS 1-specific code elements, there is no statistically significant difference between the performance of these three subgroups (Kruskal-Wallis $p = 0.94$), suggesting that the difference observed between the IPLS 1 and non-IPLS 1 groups is durable for at least two years.

To check whether taking any college physics course at all contributes positively to students' performance on IPLS 1-specific elements of the code, we also subdivided the non-IPLS 1 group into two subgroups: (i) those who had taken no college physics, and (ii) those who had taken non-IPLS college physics. We found no statistically significant difference in performance on these IPLS 1-specific elements of the code between these two groups (Wilcoxon $p = 0.87$, Fig. 1b). This result suggests that non-IPLS physics offers minimal, if any, benefit to students (relative to taking no college physics at all) on a task that relates to a physical mechanism emphasized specifically in IPLS 1.

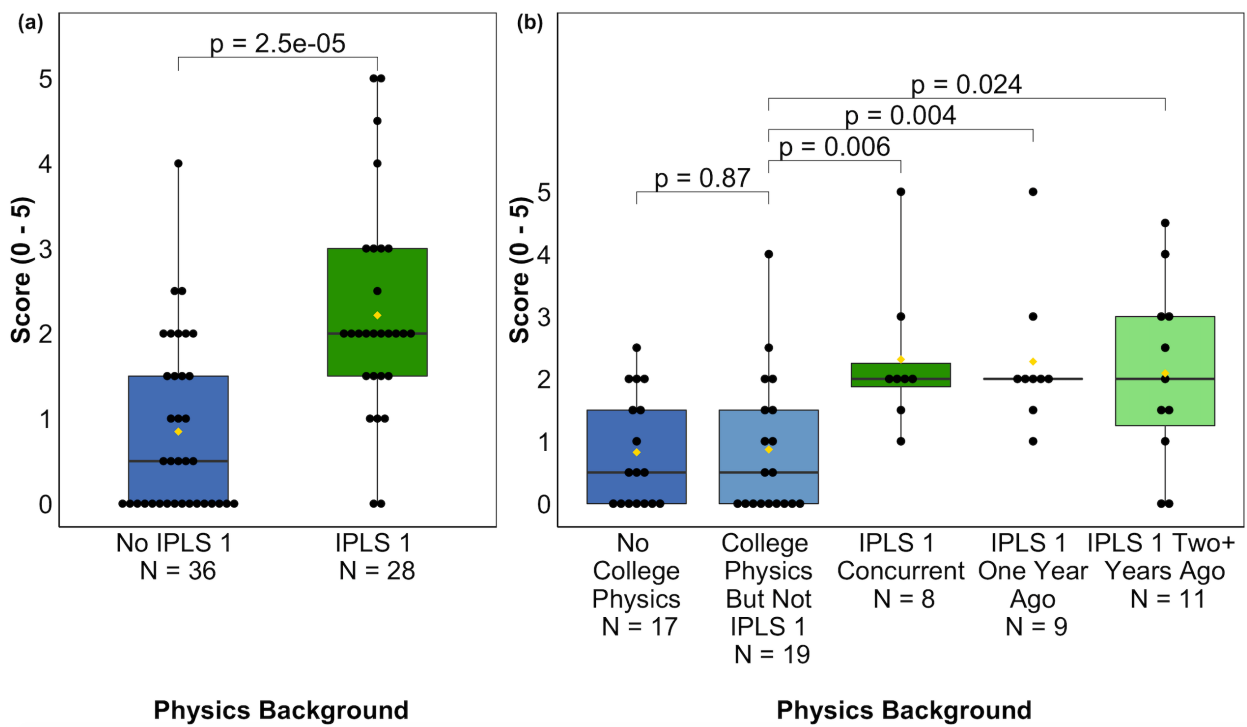

**FIG. 1.** (a) Scores on the IPLS 1-specific elements of the code for students who had and had not taken IPLS 1. Students who had taken IPLS 1 (green box plot) scored significantly higher on these elements of the code than those who had not taken IPLS (blue box plot). (b) Scores on the IPLS-1 specific elements of the code for those who had taken no college physics, those who had taken non-IPLS 1 college physics, and those who had taken IPLS 1 at different times prior to the capstone course. Taking non-IPLS physics offered minimal if any benefit to students on the IPLS 1-specific tasks (blue box plots), and the significant difference between the IPLS 1 and no-IPLS 1 groups is durable up to at least two years (green box plots). All p-values are from Wilcoxon tests.

Although it may not be surprising that IPLS 1 students scored significantly higher on elements of the code that connect directly to ideas emphasized in the IPLS 1 curriculum, we emphasize that students demonstrated these competencies in a setting far removed from the IPLS environment. Students in the biology capstone course were not explicitly primed to think about physics before completing the task, and the specific context (animal digestion) is not one that students encountered in IPLS 1. The skills and competencies that we observed were demonstrated in a *biology* course on a task administered by a *biology* instructor. Even more significantly, the ability to display these competencies is durable. The performance by students completing the task one or two years after IPLS 1 is the same as that of students completing the task concurrently with IPLS 1, within statistical variation.[2]

---

[2] Students who were concurrently taking IPLS 1 and the capstone course had completed the relevant IPLS 1 unit on diffusion well before completing the capstone task.

While IPLS 1 students score higher than their non-IPLS 1 peers on elements of the code related to IPLS 1 content and skills, *all* students found these elements challenging. The overall average on these elements of the code was 1.5 out of a possible 5 points, with the average among IPLS 1 students being slightly more than 2 out of 5. So, while the difference in performance between the IPLS 1 and non-IPLS 1 groups is significant and durable, we also find that many IPLS 1 students struggled to recall or draw upon ideas that had been discussed in IPLS 1.

### B. General Quantitative Skills

On elements of the code related to more general quantitative skills, students who completed one or more IPLS courses scored significantly higher than students who had not taken an IPLS course (Wilcoxon $p = 0.03$, Fig. 2a). Interestingly, however, this difference was not significant when we compared students who had taken only *one* semester of IPLS with those who had not taken any IPLS (Wilcoxon $p = 0.21$). The difference in performance on the general quantitative skill elements of the code was statistically significant only when we compared students who had completed both semesters of IPLS with those who had had no IPLS experience at all (Wilcoxon $p = 0.01$, Fig. 2b). This finding suggests that the development of general quantitative skills may be a cumulative process, whereby significant benefits are achieved from taking a full year of IPLS as opposed to a single semester. While the IPLS students scored higher than their non-IPLS peers on elements of the code related to general quantitative skills, Fig. 2 makes clear that students in both the IPLS and non-IPLS groups showed a wide range of scores on these code elements.

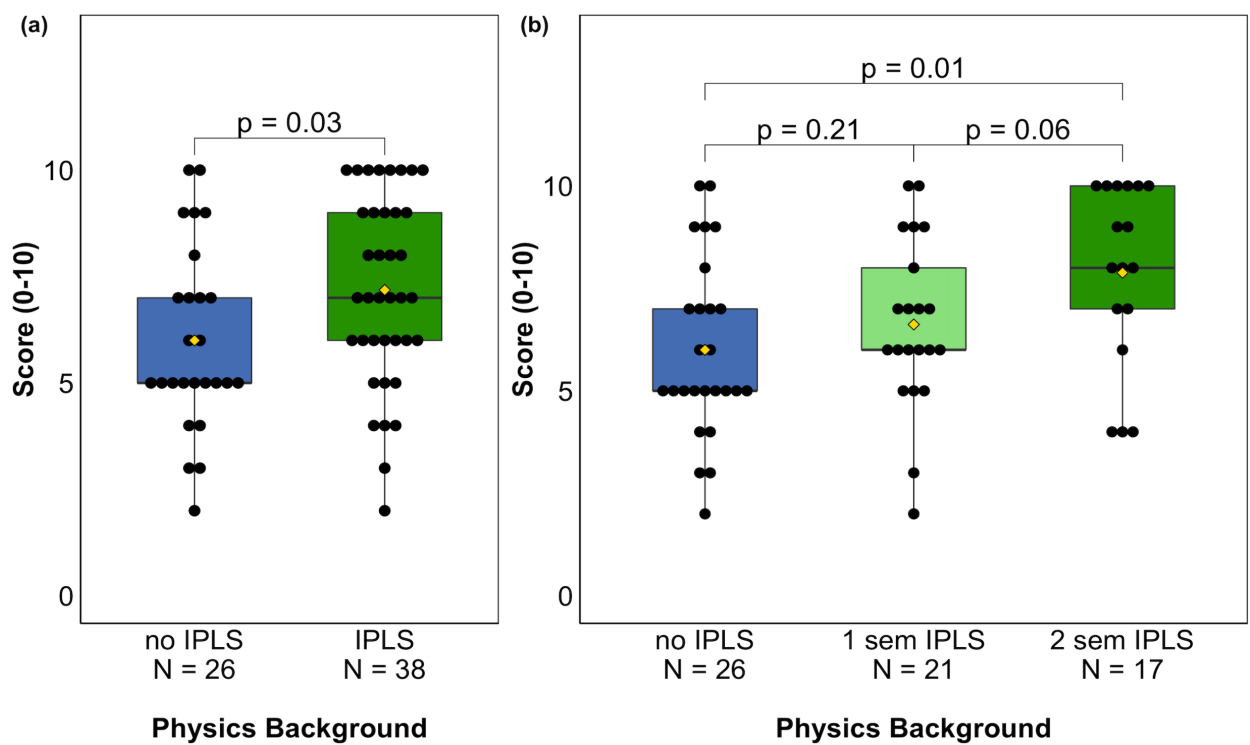

**FIG. 2.** (a) Scores on the general quantitative skill elements of the code for those who did and did not take an IPLS course. Students who took IPLS 1 (green box plot) scored significantly higher on these elements of the code than those who did not take IPLS (blue box plot). (b) Scores on the general quantitative skill elements of the code for those who had taken no IPLS course, those who had taken one IPLS course, and those who had taken both IPLS courses. Only those who had taken both IPLS courses scored significantly higher on these code elements than those who had taken no IPLS. All p-values are from Wilcoxon tests.

### C. Interpreting the Results

To assess whether the observed differences in performance on the IPLS 1-specific elements of the code (Fig. 1) and on the general quantitative skill elements of the code (Fig. 2) are due simply to IPLS students being stronger academically than their non-IPLS peers, we controlled for academic performance using an ANCOVA test with average GPA in science and engineering courses as a covariate. The ANCOVA demonstrates that scores on these task elements have minimal correlation to science course GPA, and controlling for science course GPA actually *increases* the difference between the two group means in both cases (Table IV). The difference in performance between the IPLS and non-IPLS groups, therefore, cannot be attributed to differences in overall academic performance as measured by science course GPA.

| Code elements | Observed difference in mean Δμ | Adjusted difference in mean |
|---|---|---|
| *IPLS 1-specific* | $\mu_{\text{IPLS 1}} - \mu_{\text{non-IPLS 1}} = \mathbf{1.36}$ | **1.50 (+0.14)** |
| *General quantitative* | $\mu_{\text{IPLS 1}} - \mu_{\text{non-IPLS 1}} = \mathbf{1.18}$ | **1.19 (+0.01)** |

**Table IV.** Results of ANCOVA with average Natural Science and Engineering (NSE) GPA as a covariate. For IPLS 1-specific code elements, the difference in mean between IPLS 1 and non-IPLS 1 students increases when controlling for NSE GPA (first row), and for general quantitative skill elements the difference in mean between IPLS and non-IPLS students changes negligibly when controlling for NSE GPA (second row).

To verify that performance on IPLS 1 code elements is in fact associated with having taken IPLS 1 specifically (and not just having taken *either* IPLS semester), and that performance on general quantitative skill code elements does not correlate with enrollment in a particular IPLS semester, we compared performance on the various code elements between those who had taken IPLS 1 and those who had taken only IPLS 2 (Fig. 3). Performance on the IPLS 1-specific code elements differed significantly between these groups (Wilcoxon $p < 0.001$, Fig. 3a), while performance on the general quantitative skills elements did not (Wilcoxon $p = 0.65$, Fig. 3b). This finding gives us confidence that our categorization of the codes in Table II is reasonable.

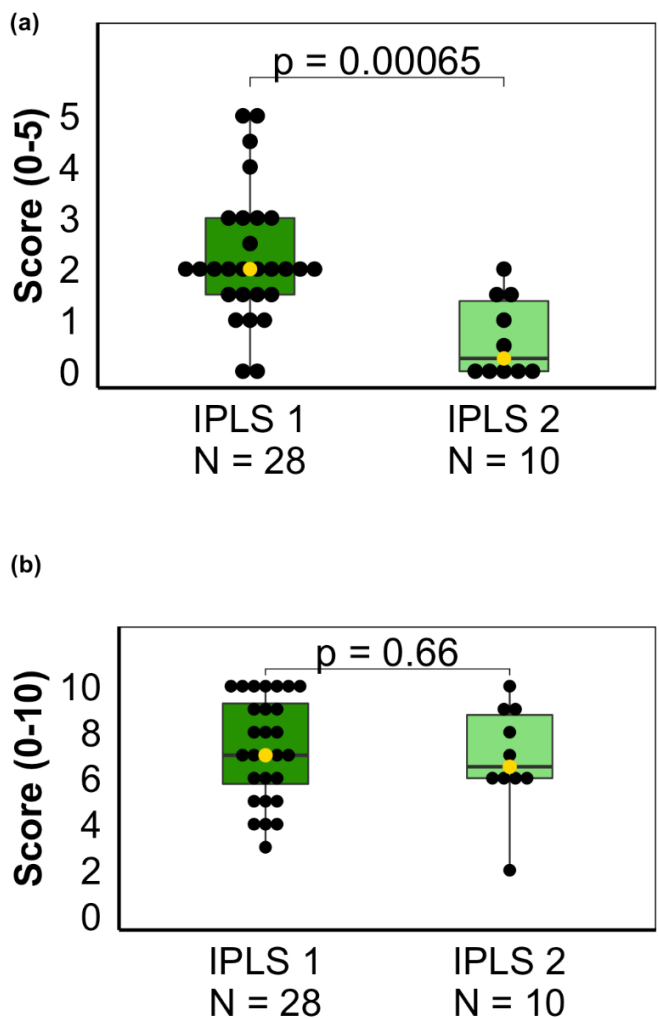

**FIG. 3.** (a) Scores on the IPLS 1-emphasized elements of the code for those who took either IPLS1 (dark green box plot) or IPLS 2 (light green box plot). Students who took IPLS 1 scored significantly higher on these elements of the code than those who took IPLS 2. (b) Scores on the general quantitative skill elements of the code for those who took either IPLS 1 (dark green box plot) or IPLS 2 (light green box plot). There is no significant difference between the two groups on these elements of the code. All p-values are from Wilcoxon tests.

IPLS 1 students scored higher on IPLS 1-specific elements of the code (Fig. 1), and IPLS students (especially those taking both semesters of IPLS) score higher on the general quantitative skill elements of the code (Fig. 2). However, *all* of the students completing the capstone task in our study, whether they had taken IPLS or not, had taken General Chemistry and two Math or Statistics courses, and most students had taken two additional quantitative courses from the set of courses eligible to meet the quantitative requirement, including Organic Chemistry, Introductory

Computer Science, Introductory Physics, Introductory Computer Science, Introductory Engineering, and others. These courses are designated as satisfying the department quantitative course requirements precisely because they afford students the opportunity to develop quantitative skills. We were interested, therefore, in investigating whether some other course(s) among those that satisfy the biology department requirements might also lead to the kinds of differences in performance that we observed. Because everyone in the study had taken General Chemistry and two semesters of Introductory Math or Statistics, we could not use those courses for a meaningful comparison. Instead, we chose to compare performance on the diffusion task between students who had or had not taken at least one course in computer science (CS). While computer science courses count toward the quantitative course requirements in the biology department, our hypothesis was that the particular skills developed in computer science courses would not map cleanly onto the skills being assessed in the capstone task, and therefore we would not see a similar result. Indeed, we found no significant difference between the CS and non-CS groups on quantitative skills elements of the code (Wilcoxon $p = 0.62$).

While this finding does not rule out the possibility that some *other* course that students are taking is responsible for the differences in performance on IPLS 1-specific and general quantitative elements of the code, it does demonstrate that the differences we observe are not due to students taking just *any* course with quantitative skill building elements. The consistent coordination between complex biological phenomena and simple physical models that is the hallmark of the Swarthmore IPLS courses indeed appears to support students in applying physical reasoning to the diffusion task.

## VII. CONCLUSIONS AND OUTLOOK

This article presents findings from a longitudinal interdisciplinary study in which we compared reasoning exhibited by IPLS and non-IPLS students on a diffusion task administered at the end of a biology senior capstone course. We found that IPLS 1 students were more likely than non-IPLS 1 students to reason mechanistically about diffusive phenomena and to successfully coordinate between multiple representations of diffusive processes, even up to two years after taking the IPLS 1 course. IPLS students also out-performed their non-IPLS peers on metrics related to general quantitative reasoning. Further, these IPLS-derived gains appear to be cumulative, with two semesters of IPLS offering greater benefit than just one semester.

As described in Part A of Section IV, the work reported here comes after several years of previous efforts to assess the long-term outcomes of our IPLS course at Swarthmore. We have observed that our ability to measure these outcomes, in particular the demonstration of physical reasoning on biological tasks administered in biology classes, depends on the degree to which we as IPLS researchers have input into the design of the assessment task. Further work must be done to determine how best to elicit physical reasoning from students in settings near and far from the IPLS environment, in order to effectively tease out what students have and have not retained from their IPLS experience.

Finally, we note that we have not answered in this article the essential question of *how* the curricular and pedagogical features of the IPLS environment support students in developing and applying physical and quantitative skills. As described in Part I, the IPLS course involves explicit messaging about the connections between physics and other disciplines, provides students with multiple opportunities to practice making such connections, and supports students in seeing themselves as capable of applying physics in novel settings. All of these factors interact to create the IPLS ecosystem and to support students in applying what they have learned beyond the boundaries of the IPLS classroom. Further work is required to more precisely model the relationship between these factors and the durability of the attitudes and skills that we observe.

## ACKNOWLEDGMENTS

We gratefully acknowledge Michelle Smith for suggesting the biology capstone course as an environment for assessment, and Brad Davidson for working with us to administer the task during the second year of the capstone course. We thank Brandon Daniel-Morales, Nikhil Tignor, and Calvin White for producing the figures, and Nathaniel Peters and Jonathan Solomon for analyzing prior student work on embedded tasks. Finally, we thank our longitudinal study advisory board —Eric Brewe, Brad Davidson, Todd Cooke, Eric Kuo, and Sanjay Rebello — for providing valuable feedback and suggestions at several stages. This work was funded by the National Science Foundation under the IUSE program (EHR-1710875).

[1] Howard Hughes Medical Institute-American Association of Medical Colleges Committee, *Scientific Foundations for Future Physicians*, Am. Assoc. Medial Coll. 46 (2009).

[2] American Association for the Advancement of Science, *Vision and Change in Undergraduate Biology Education* (2009).

[3] National Research Council, *BIO2010: Transforming Undergraduate Education for Future Research Biologists* (The National Academies Press, Washington, DC, 2003).

[4] National Research Council, *A New Biology for the 21st Century* (National Academies Press, 2009).

[5] E. F. Redish, C. Bauer, K. L. Carleton, T. J. Cooke, M. Cooper, C. H. Crouch, B. W. Dreyfus, B. D. Geller, J. Giannini, J. S. Gouvea, M. W. Klymkowsky, W. Losert, K. Moore, J. Presson, V. Sawtelle, K. V. Thompson, C. Turpen, and R. K. P. Zia, *NEXUS/Physics: An Interdisciplinary Repurposing of Physics for Biologists*, Am. J. Phys. **82**, 368 (2014).

[6] C. H. Crouch and K. Heller, *Introductory Physics in Biological Context: An Approach to Improve Introductory Physics for Life Science Students*, Am. J. Phys. **82**, 378 (2014).

[7] D. C. Meredith and E. F. Redish, *Reinventing Physics for Life-Sciences Majors*, Phys. Today **66**, 38 (2013).

[8] E. F. Redish and T. J. Cooke, *Learning Each Other's Ropes: Negotiating Interdisciplinary Authenticity*, CBE Life Sci. Educ. **12**, 175 (2013).

[9] B. W. Dreyfus, V. Sawtelle, C. Turpen, J. Gouvea, and E. F. Redish, *Students' Reasoning about "High-Energy Bonds" and ATP: A Vision of Interdisciplinary Education*, Phys. Rev. Spec. Top. - Phys. Educ. Res. **10**, 010115 (2014).

[10] B. D. Geller, J. Gouvea, B. W. Dreyfus, V. Sawtelle, C. Turpen, and E. F. Redish, *Bridging the Gaps: How Students Seek Disciplinary Coherence in Introductory Physics for Life Science*, Phys. Rev. Phys. Educ. Res. **15**, 020142 (2019).

[11] C. H. Crouch, P. Wisittanawat, M. Cai, and K. A. Renninger, *Life Science Students' Attitudes, Interest, and Performance in Introductory Physics for Life Sciences: An Exploratory Study*, Phys. Rev. Phys. Educ. Res. **14**, 010111 (2018).

[12] B. D. Geller, C. Turpen, and C. H. Crouch, *Sources of Student Engagement in Introductory Physics for Life Sciences*, Phys. Rev. Phys. Educ. Res. **14**, 010118 (2018).

[13] W. Bialek and D. Botstein, *Introductory Science and Mathematics Education for 21st-Century Biologists.*, Science **303**, 788 (2004).

[14] J. Watkins, J. E. Coffey, E. F. Redish, and T. J. Cooke, *Disciplinary Authenticity: Enriching the Reforms of Introductory Physics Courses for Life-Science Students*, Phys. Rev. Spec. Top. - Phys. Educ. Res. **8**, 010112 (2012).

[15] J. Watkins and A. Elby, *Context Dependence of Students' Views about the Role of Equations in Understanding Biology*, CBE—Life Sci. Educ. **12**, 274 (2013).

[16] K. L. Hall, *Examining the Effects of Students' Classroom Expectations on Undergraduate Biology Course Reform* (University of Maryland, College Park 2013).

[17] V. Sawtelle and C. Turpen, *Leveraging a Relationship with Biology to Expand a Relationship with Physics*, Phys. Rev. Spec. Top. - Phys. Educ. Res. **12** (2016).

[18] M. M. Cooper and M. W. Klymkowsky, *The Trouble with Chemical E4nergy: Why Understanding Bond Energies Requires an Interdisciplinary Systems Approach.*, CBE Life Sci. Educ. **12**, 306 (2013).

[19] J. S. Gouvea, V. Sawtelle, B. D. Geller, and C. Turpen, *A Framework for Analyzing Interdisciplinary Tasks: Implications for Student Learning and Curricular Design*, CBE—Life Sci. Educ. **12**, 187 (2013).

[20] B. D. Geller, B. W. Dreyfus, J. Gouvea, V. Sawtelle, C. Turpen, and E. F. Redish, *Entropy and Spontaneity in an Introductory Physics Course for Life Science Students*, Am. J. Phys. **82**, 394 (2014).

[21] B. W. Dreyfus, J. Gouvea, B. D. Geller, V. Sawtelle, C. Turpen, and E. F. Redish, *Chemical Energy in an Introductory Physics Course for the Life Sciences*, Am. J. Phys. **82**, 403 (2014).

[22] R. A. Engle, *Framing Interactions to Foster Generative Learning: A Situative Explanation of Transfer in a Community of Learners Classroom.*, J. Learn. Sci. **15**, 451 (2006).

[23] R. A. Engle, P. D. Nguyen, and A. Mendelson, *The Influence of Framing on Transfer: Initial Evidence from a Tutoring Experiment.*, Instr. Sci. **39**, 603 (2011).

[24] R. A. Engle, D. P. Lam, X. S. Meyer, and S. E. Nix, *How Does Expansive Framing Promote Transfer? Several Proposed Explanations and a Research Agenda for Investigating Them*, Educ. Psychol. **47**, 215 (2012).

[25] A. Collins, J. S. Brown, and A. Holum, *Cognitive Apprenticeship: Making Thinking Visible*, American Educator, **15** (1991).

[26] P. W. Irving, D. McPadden, and M. D. Caballero, *Communities of Practice as a Curriculum Design Theory in an Introductory Physics Class for Engineers*, Phys. Rev. Phys. Educ. Res. **16**, 020143 (2020).

[27] J. D. Bransford and D. L. Schwartz, *Rethinking Transfer: A Simple Proposal with Multiple Implications*, Rev. Res. Educ. **24**, 61 (1999).

[28] J. Bransford, National Research Council (U.S.) (Eds.), *How People Learn: Brain, Mind, Experience, and School*, Expanded ed (National Academy Press, Washington, D.C., 2000).

[29] G. Rak, B. D. Geller, and C. H. Crouch, *Assessing the Lasting Impact of IPLS on Student Interdisciplinary Attitudes*, contributed talk at the virtual Physics Education Research Conference (2020).

[30] J. Solomon, N. Peters, B. D. Geller, C. Turpen, and C. H. Crouch, *Assessing the Lasting Impact of an IPLS Course*, contributed talk at the Physics Education Research Conference, Washington D.C. (2018).

[31] J. Rubien, Crouch, Catherine H., Hiebert, Sara M., and Geller, Benjamin D., *The Impact of IPLS in a Senior Biology Capstone Course*, contributed talk at the virtual Physics Education Research Conference (2020).

[32] Tipton, Maya, Crouch, Catherine H., and Geller, Benjamin D., *Does IPLS Help Students Apply Physics to Biology?*, contributed talk at the virtual Physics Education Research Conference (2020).

[33] The Concord Consortium, *Diffusion and Temperature Activity*, Concord.org (2021).

[34] Trampleasure, L. *Diffusion of food coloring in hot and cold water*, YouTube.com (2020).

[35] McDonald, J.H., *Handbook of Biological Statistics (3rd Ed.)* (Sparky House Publishing, Baltimore, MD, 2014).

[36] T. J. VanderWeele and M. B. Mathur, *Some Desirable Properties of the Bonferroni Correction: Is the Bonferroni Correction Really so Bad?*, Am. J. Epidemiol. **188**, 617 (2019).

[37] J. Green and N. Britten, *Qualitative Research and Evidence Based Medicine*, BMJ **316**, 1230 (1998).

**APPENDIX** (Biology capstone diffusion task)

NAME: ______________________________

## SCENARIO DESCRIPTION (read carefully)

An animal consumes a large, very fatty meal. Fatty acids from this meal enter the blood quickly, following this path:

First, fatty acids diffuse freely across the cell membrane to enter one of the cells lining the intestinal wall. They then diffuse across the cell, across the cell membrane on the other side of this cell, and finally into the blood vessel, where the flow of blood immediately carries the fatty acids away from the intestinal wall.

***Assume that the supply of fatty acids in the fatty meal is unlimited, and that the blood supply is unlimited is well.***

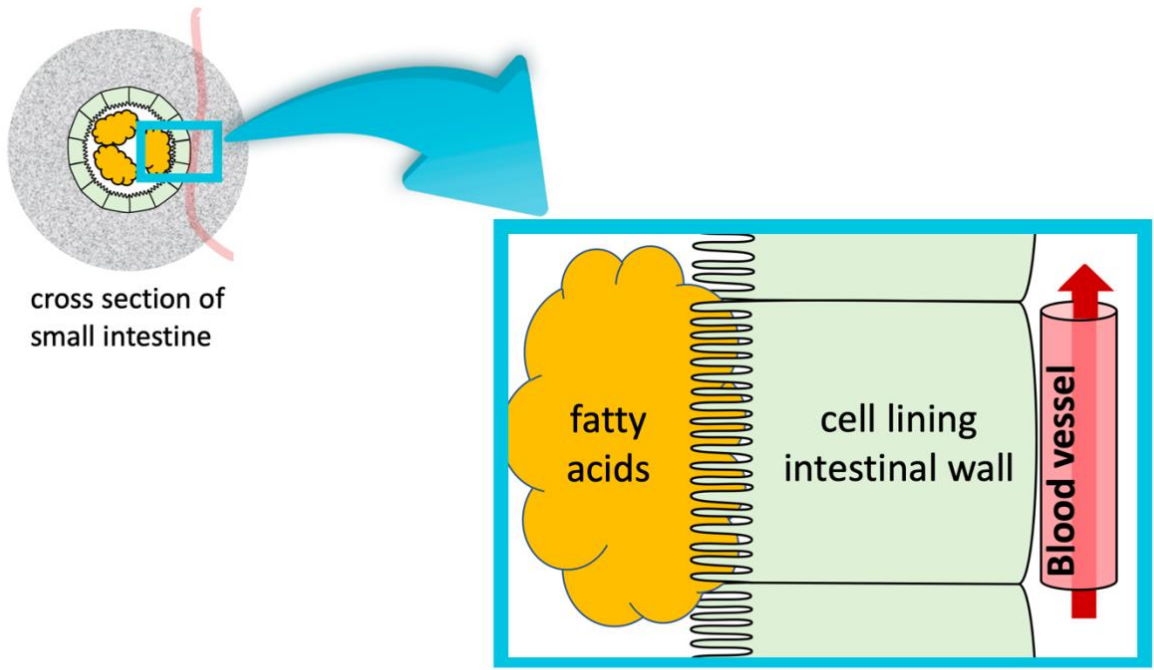

NAME: ______________________________

## Part I.

1. **Draw a plot of the fatty acid concentration $c$ (molecules/$\mu m^3$) as a function of position $x$ ($\mu m$), using the axes labeled below. Annotate your plot with short descriptions that indicate why you have drawn things as you have.** Your plot should show how the concentration varies with position in all three regions (the small intestine, the cell, and the blood vessel). You may assume that the concentration of fatty acids decreases linearly as you move from the small intestine to the blood vessel across the cell lining the intestinal wall. Remember that there is unlimited supply of fatty acids from the food and unlimited blood supply.

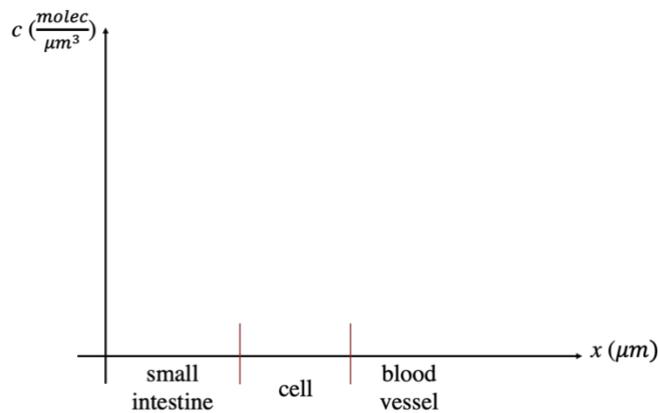

2. Any individual fatty acid molecule moves randomly, like the red molecule in the simulation we just watched. That is, in any moment the fatty acid molecule is just as likely to move left or right. However, we also know that fatty acid molecules do end up migrating from the small intestine to the blood vessel via diffusion! Explain this apparent paradox: **Why is there net movement of fatty acids from left to right in the figure even though any individual fatty acid molecule is equally likely to move left or right?** Be as precise as possible in your explanation, and feel free to include a figure or drawing if that helps. You can continue your response on the back of this page.

NAME: _______________________________

## Part II.

1. Consider the four plots A, B, C, and D shown below. In each case, fatty acid concentration $c$ is plotted as a function of position $x$. A cartoon showing the situation accompanies each plot. As the cartoons and plots illustrate, the size of the cell lining the intestinal wall is not the same in every plot, and the initial fatty acid concentration in the small intestine is not the same in every plot.

A. 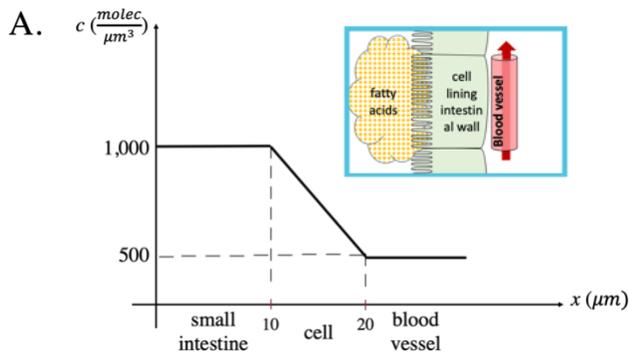

B. 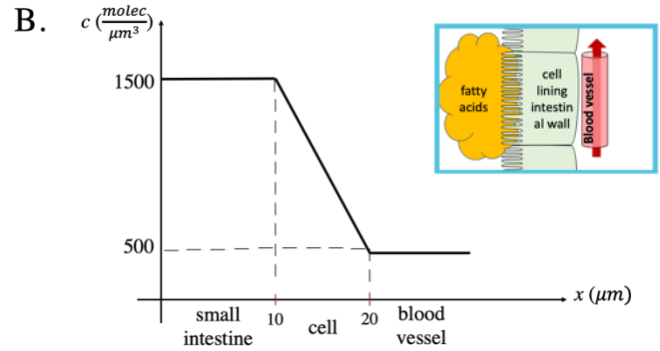

C. 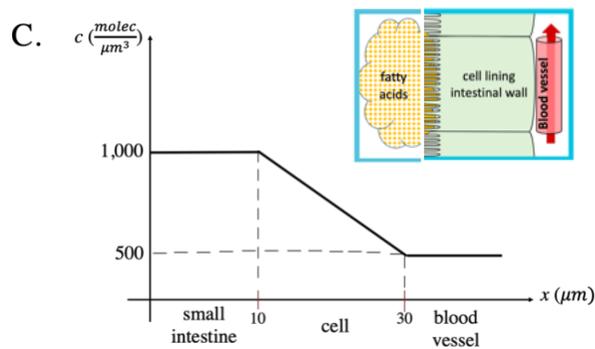

D. 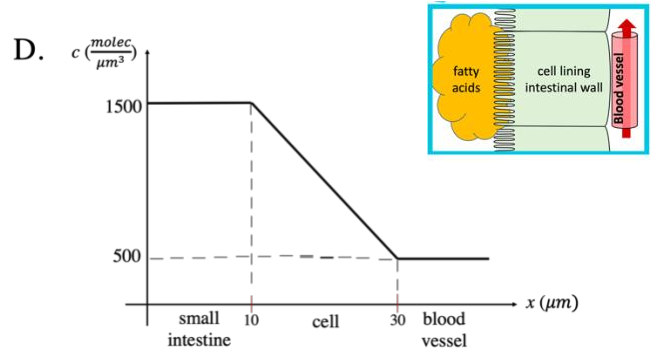

**In the box at right, rank the <u>rates</u> at which fatty acids diffuse from the small intestine to the blood vessel in the four cases (A, B, C, D), from fastest to slowest:**

fastest                                  slowest

Carefully explain how you arrived at your ranking. If two rates are the same, explain how you know that. Feel free to continue your response on the back.

NAME: _______________________________

## Part III.

You may have heard of Fick's Law, which describes the rate at which molecules diffuse along a concentration gradient. When two regions (regions 1 and 2 in the figure below) differ in molecular concentration, the rate at which molecules diffuse from region 1 to region 2 is proportional to the difference in concentration between the two regions $\Delta c$, and inversely proportional to the distance separating the two regions $\Delta x$.

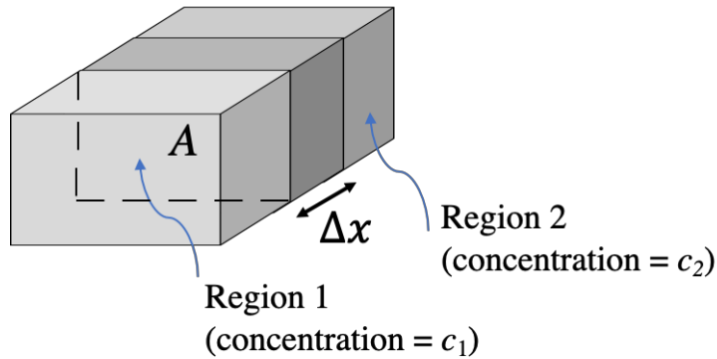

Assume that the figure above illustrates a situation where the concentration in each of regions 1 and 2 is constant (but different in each region). You may also assume for simplicity that the concentration changes linearly as one moves from region 1 to region 2 across the middle (darkest) region that has length $\Delta x$. The cross-sectional area between the two regions is $A$, as shown.

The proportionality constant between the rate of diffusion per cross-sectional area and the ratio $\Delta c/\Delta x$ is defined as the diffusion constant $D$. The value of $D$ varies from molecule to molecule, and depends also on the medium in which the molecules are diffusing.

Fick's Law allows us to mathematically represent the situation just described:

$$J = -D\frac{\Delta c}{\Delta x}$$

In this equation:

- $J$ = the rate at which molecules diffuse per cross sectional area (unit = molecules/m²*s)
- $D$ = the diffusion constant for the particular molecule at a particular temperature in the particular medium through which it is diffusing (unit = m²/s)
- $\Delta c$ = the molecular concentration difference between the two regions (unit = molecules/m³)
- $\Delta x$ is the distance separating the two regions (unit = m)

1. You may have seen Fick's Law written without a negative sign, because in some cases you are only interested in absolute values of diffusion rates. However, the minus sign is important. **Why is the minus sign essential to a complete statement of Fick's Law?** Explain your reasoning carefully, and feel free to include a drawing or figure if it helps with your explanation.

2. Consider the plot of fatty acid concentration as a function of position shown below. The diffusion constant for fatty acids in this context is $D = 100 \ \mu m^2/s$. Numerical concentration and position values are provided along the axes. **At each of the labeled locations (A, B, and C), calculate the number of fatty acid molecules that move per second through a cross-sectional patch of area $A = 2 \ \mu m^2$. The unit of your answer in each region should be molecules/s, and you should <u>comment on the meaning of the signs</u> of your answers.**

A:

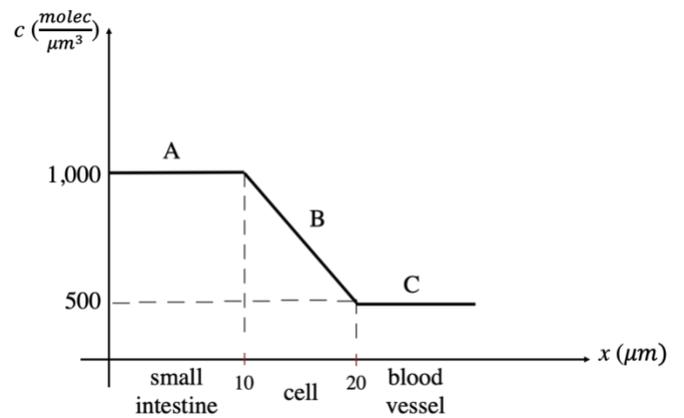

B:

C: